\documentstyle[12pt]{article}
\begin{document}
\baselineskip=21pt
\rightline{YCTP-P15-92}
\rightline{April 1992}
\vskip .2in
\begin{center}
{\large{\bf Grand Unification, Gravitational Waves, and
the Cosmic Microwave Background Anisotropy}}
\end{center}
\vskip .1in
\begin{center}
Lawrence M.
Krauss\footnote{Also Dept. of Astronomy.  Research supported in part
by the NSF,DOE, and TNRLC}
and Martin White\footnote{Address after September 1, Center for Particle
Astrophysics, Berkeley}

{\it Center for Theoretical Physics, Sloane
Laboratory}

{\it  Yale University, New Haven, CT 06511}
\end{center}

\vskip .2in

\centerline{ {\bf Abstract} }

\baselineskip=18pt

\noindent
We re-examine the gravitational wave background resulting from inflation and
its effect on the cosmic microwave background radiation. The new COBE
measurement of a cosmic background quadrupole anisotropy places an upper limit
on the vacuum energy during inflation of $\approx 5 \times 10^{16}$ GeV.
A stochastic background of gravitational waves from inflation could produce
the entire observed signal (consistent with the observed dipole anisotropy and
a flat spectrum) if the vacuum energy during inflation was as small as
$1.5 \times 10^{16}$ GeV at the 95\% confidence level.
This coincides nicely with the mass scale for Grand Unification inferred from
precision measurements of the electroweak and strong coupling constants, for
the SUSY Grand Unified Theories.  Thus COBE could be providing the first direct
evidence, via gravitational waves, for GUTs, and supersymmetry. Further tests
of this possibility are examined, based on analyzing the energy density
associated with gravitational waves from inflation.

\baselineskip=21pt

\newpage

The observation by the DMR instrument aboard the COBE satellite of
large scale anisotropies in the Cosmic Microwave
background (CMB) \cite{smoot} is probably the most important discovery in
cosmology since the discovery of the CMB itself \cite{penzias}.  Such
anisotropies cannot have been induced by causal processes which were
initiated after the era of recombination and thus represent true
primordial fluctuations resulting from physics associated with the
initial conditions of the FRW cosmology itself.  These initial
conditions are likely to have resulted from processes associated with
either an inflationary phase or new planck scale physics.  Only in the former
case can explicit predictions be made and the COBE data on the temperature
correlation function is remarkably consistent with a flat Harrison-Zel'dovich
spectrum as predicted from inflation.

Inflation predicts at least two sources of Harrison-Zel'dovich type CMB
anisotropies.  Scalar energy density fluctuations on the surface of last
scattering induced by primordial (dark) matter density perturbations will
result both in subsequent structure formation and in appropriate dipole,
quadrupole, and higher moment anisotropies in the CMB \cite{gys}. Based on the
observed dipole asymmetry one can determine an upper limit on the expected
quadrupole anisotropy in the case of such a flat spectrum.  In addition, if
the scale of inflation is sufficiently high, long wavelength gravitational
waves will be generated during inflation whose re-entry into the horizon can
result in a large scale observed quadrupole and higher multipole anisotropies
in the CMB today \cite{gw}. While inflation is not the only method of
generating a stochastic background of horizon-sized waves \cite{krauss}, it
is certainly the most well motivated.

In this work we re-examine gravitational wave generation during inflation and
determine the predicted signal in the CMB and compare this with the COBE data.
In the process of deriving detailed estimates we update and reconcile various
earlier analyses.  We present a likelihood function for the probability that
inflation at a given scale would result in a quadrupole anisotropy at least as
big as that which is observed.  We also briefly compare this to the nature and
magnitude of the expected quadrupole anisotropy resulting from scalar density
perturbations expected from inflation.  Our analysis allows us to place both
upper and lower limits on the range of scales for which gravitational waves
from inflation could result in all or most of the observed quadrupole
anisotropy.  These scales are consistent with the scale at which the
$SU(3)\times SU(2)\times U(1)$ gauge couplings can be unified, based on a
renormalization group extrapolation of low energy data, for SUSY GUT models.
We find this coincidence both suggestive and exciting, and consider other
possible observational probes of a gravitational wave background at this level.
For this purpose we calculate the energy density stored in such a stochastic
background today.

\vskip .1in

Since the work of Starobinsky \cite{star}, it has been recognized that a
period of exponential expansion in the early universe would lead to the
production of gravitational waves.
Rubakov and collaborators \cite{Anis} were the first to use this to limit the
scale of inflation and with it the scale of Grand Unification.  Since that
time analyses designed to help more accurately compute the gravitational wave
background and compare predictions to the data have been developed
\cite{FabPol,AbbWis1,AbbSch,Sahni}. More recently the the limits on the
quadrupole anisotropy of the microwave background had improved.  It thus
seemed, even before COBE, a propitious time to re-analyze the gravitational
wave limits.  Many of the analytic techniques and results we derive have
appeared in one form or another scattered in the literature, but we have made
some effort to check, unify and reconcile the previous methods and in the
process correct any errors.  Further details can be found in \cite{white}.

It is convenient to write the metric in the $k=0$ Robertson-Walker form
\begin{equation}
  ds^2 =  R^2(\tau)\left( -d\tau^2 + d\vec{x}^2 \right)
\label{eqn:metric}
\end{equation}
where $d\tau = dt/R(t)$ is the conformal time.
In a universe which undergoes a period of exponential inflation, followed by a
radiation dominated epoch and then a matter dominated phase, $R(\tau)$ and
$\dot{R}(\tau)$ can be matched at the transition points, assuming that the
transitions between phases are sudden  (this approximation is sufficient for
our purposes) \cite{AbbHar}.  Because of the matching conditions $\tau$ is
discontinuous across the transitions.  We define $\tau_1$ to be the (conformal)
time of radiation-matter equality, $\tau_2$ to be the end of inflation.
The Hubble constant during inflation, $H$, and vacuum energy density $V_0$
driving the inflation are related by
\begin{equation}
  H^2 = {8\pi\over 3} {V_0\over m_{Pl}^2} = {8\pi\over 3} m_{Pl}^2 v
\label{eqn:hubble-v0}
\end{equation}
where we introduce the notation $v\equiv V_0/m_{Pl}^4$.

A classical gravitational wave in the linearized theory is a ripple on the
background space-time
\begin{equation}
  g_{\mu\nu} = R^2(\tau) \left( \eta_{\mu\nu} + h_{\mu\nu} \right)
  \ \ \mbox{where }\ \eta_{\mu\nu}=\mbox{diag}(-1,1,1,1),
  \ \ h_{\mu\nu} \ll 1
\label{eqn:wave-metric}
\end{equation}
In transverse traceless (TT) gauge the two independent polarization states of
the wave are denoted as $+,\times$.  In the linear theory the TT metric
fluctuations are gauge invariant (they can be related to components of the
curvature tensor).  Write
\begin{equation}
  h_{\mu\nu}(\tau,\vec{x}) = h_{\lambda}(\tau;\vec{k})
  e^{i\vec{k}\cdot\vec{x}} \epsilon_{\mu\nu}(\vec{k};\lambda)
\label{eqn:plane-wave}
\end{equation}
where $\epsilon_{\mu\nu}(\vec{k};\lambda)$ is the polarization tensor and
$\lambda=+,\times$.  The equation for the amplitude
$h_{\lambda}(\tau;\vec{k})$ is obtained by requiring the perturbed metric
(\ref{eqn:wave-metric}) satisfy Einstein's equations to $O(h)$.
As was first noted by Grishchuk \cite{Grishchuk} the equation of motion for
this amplitude is then identical to the massless Klein-Gordon equation for a
plane wave in the background space-time.  In this way, one finds each
polarization state of the wave behaves as a massless, minimally coupled, real
scalar field, with a normalization factor of $\sqrt{16\pi G}$ relating the two.

The spectrum of gravitational waves generated by quantum fluctuations during
the inflationary period can be derived by a sequence of Bogoluibov
transformations relating creation and annihilation operators defined in the
various phases: inflationary, radiation and matter dominated
\cite{AbbHar,Sahni}.  The key idea is that for long wavelength (c.f. the
horizon size) modes the transitions between the phases are sudden and the
universe will remain in the quantum state it occupied before the transition
(treating each of the transitions as instantaneous is a good approximation for
all but the highest frequency graviton modes).  However the creation and
annihilation operators that describe the particles in the state are related
by a Bogoluibov transformation, so the quantum expectation value of any
string of fields is changed.
A calculation of the quantum $n$-point functions suffices to find the spectrum
of classical gravitational waves today since the statistical average of the
ensemble of classical waves can be related to the corresponding quantum
average.

A stochastic spectrum of classical gravitational waves (in terms
of comoving wavenumber $\vec{k}$) in the expanding universe
 has the form
\begin{equation}
  h_{\lambda}(\tau ;\vec{k}) = A(k) a_{\lambda}(\vec{k})
  \left[{3j_1(k\tau)\over k\tau}\right]\ \ \ \mbox{with}\ \ \lambda=+,\times
\label{eqn:spectral-amp}
\end{equation}
where the term $[\cdots]$ is a real solution of the Klein-Gordon
equation in a matter dominated universe.  $a_{\lambda}(\vec{k})$ is a random
variable with statistical expectation value (normalized to simplify the
relation between $A(k)$ and the energy density per logarithmic frequency
interval as we later describe)  \begin{equation} \left\langle
a_{\lambda}(\vec{k}) a_{\lambda'}(\vec{q}) \right\rangle = k^{-3}
\delta^{(3)}(\vec{k}-\vec{q}) \delta_{\lambda\lambda'} \label{eqn:rv-norm}
\end{equation}

Waves which are still well outside the horizon at the time of matter-radiation
equality $(k\tau_1\ll 2\pi)$ will give the largest contribution to the CMB
anisotropy today. Calculating the Bogoluibov coefficients by matching the
field and its first derivative at $\tau_2,\tau_1$ in the limit $k\tau\ll 2\pi$
one can derive the prediction for the ($k$-independent) spectrum of
long-wavelength gravitational waves generated by inflation
\cite{gw,white}
\begin{equation}
A^2(k) = {H^2\over\pi^2 m_{Pl}^2} = {8\over 3\pi}v
\label{eqn:amplitude}
\end{equation}

To make contact with observations one must consider the effect such a
spectrum will have on the CMBR.
If one expands the CMBR temperature anisotropy in spherical harmonics
\begin{equation}
{\delta T\over T}(\theta,\phi) = \sum_{lm} a_{lm} Y_{lm}(\theta,\phi)
\label{eqn:mode-expansion}
\end{equation}
one can present the prediction of a given spectrum of gravitational waves
in terms of the $a_{lm}$.
The temperature fluctuation due to a gravitational wave $h_{\mu\nu}$ can be
found using the Sachs-Wolfe formula
\begin{equation}
  {\delta T\over T} = -{1\over 2} \int_e^r d\Lambda
  \ {\partial h_{\mu\nu}(\tau,\vec{x})\over \partial\tau}
  \hat{x}^{\mu} \hat{x}^{\nu}
\label{eqn:sachs-wolfe}
\end{equation}
where $\Lambda$ is a parameter along the unperturbed path and the lower (upper)
limit of integration represents the point of emission (reception) of the
photon.

It is standard to project out a multipole and calculate the rotationally
symmetric quantity
\begin{equation}
  \left\langle a_l^2 \right\rangle \equiv
  \left\langle \sum_m |a_{lm}|^2 \right\rangle
\end{equation}
After some algebra, utilizing identities for spherical polynomials and Bessel
functions (i.e. see \cite{white}), one finds for waves entering the horizon
during the matter dominated era (the results are very insensitive to this
restriction since the $k$-integral is dominated by waves entering the horizon
today: $k\approx 2\pi /\tau_0$)
\begin{equation}
\left\langle a_l^2 \right\rangle = 36\pi^2 (2l+1){(l+2)!\over (l-2)!}
\int_0^{2\pi/\tau_1} k dk\ A^2(k) | F_l(k) |^2
\label{eqn:moment}
\end{equation}
where the function $F_l(k)$ is defined as ($\tau(r)=\tau_0-r$)
\begin{equation}
F_l(k) \equiv
\int_0^{\tau_0-\tau_1} dr \left({d\over d(k\tau)}{j_1(k\tau)\over k\tau}\right)
\left[ {j_{l-2}(kr)\over(2l-1)(2l+1)}+{2j_l(kr)\over(2l-1)(2l+3)}
      +{j_{l+2}(kr)\over(2l+1)(2l+3)} \right]
\end{equation}
Accounting for the factor of two difference between definitions of $A^2(k)$
this agrees with the result of \cite{AbbWis1}, and differs by $\approx 2$
with the earlier result of \cite{FabPol}.

The calculation of the expectation value $\langle a_l^2 \rangle$ is not the
end of the story however. One must also consider the statistical properties of
$a_l^2$ \cite{FabPol,AbbWis2,bond,white}.  Given that each of the $a_{lm}$ are
independent Gaussian random variables the probability distribution for each
$a_l^2$, with mean $\langle a_l^2\rangle$, is of the $\chi^2$ form. One can
calculate the confidence levels for $a_l^2$ in terms of the incomplete gamma
function. We find for the for the quadrupole, ${\langle a_2^2\rangle}=7.74v$
and $a_2^2/\langle a_2^2\rangle =$ .63,.32,.23 at the 68, 90 and 95\% (lower)
confidence levels respectively.

The new COBE observations can be summarized for our purposes as as a value for
the rms quadrupole moment.  If one fits to a Harrison-Zel'dovich spectrum the
quoted value is \cite{smoot}
\begin{equation}
Q_{rms-PS} \equiv \left({ a_2^2 \over 4\pi}\right)^{1/2} =
{(16.7\pm 4)\mu K\over 2.73K} \Rightarrow  a_2^2 = (4.7\pm 2) \times 10^{-10}.
\label{eqn:qrms}
\end{equation}
Note that the quoted error on $Q_{rms-PS}$ is Gaussian, while the distribution
of $a_2^2$ is $\chi^2$.  This implies that in proceeding from the inferred
value of $a_2^2$ to a value of $v$ we must be careful to properly take into
account the resultant statistics which will be far from Gaussian. In particular
the mode of the distribution will be lower than the mean (as is noted in
\cite{smoot}).  From (\ref{eqn:qrms}) the quadrupole moment is consistent with
gravitational waves resulting from a mean value of $v = 6.1 \times 10^{-11}$.
To determine the uncertainty on $v$ we have performed a simple Monte-Carlo
analysis to find the distribution for $v$ (see figure 1).
Based on this we can determine both upper and lower limits on the value of $v$
consistent with the observations and also find the most probable value of $v$.
We find
\begin{equation}
\begin{array}{cccccc}
 3.7 \times 10^{-10} & > & v & > & 2.5 \times 10^{-12} & 95\%\  CL \\
 1.5 \times 10^{-10} & > & v & > & 2.3 \times 10^{-11} & 68\%\  CL
\end{array}
\label{eqn:limits}
\end{equation}
with a maximum likelihood value of $v \approx 4 \times 10^{-11}$.

These limits as quoted require some interpretation.  First the 95\% upper
limit $v<3.7\times 10^{-10}$ provides a strict upper limit on the scale of
inflation $\approx (v^{1/4}M_{Pl})=5.2\times 10^{16}$GeV assuming that the
contribution to the quadrupole moment from scalar density perturbations is
insignificant.  This could be increased slightly in the unlikely case that a
comparable quadrupole moment from density perturbations existed and happened
to cancel out that due to gravitational waves to some degree.

In this regard it is worthwhile considering what magnitude of quadrupole moment
is expected from scalar density perturbations from inflation. By requiring
that the induced dipole due to long wavelength modes not greatly exceed the
observed dipole anisotropy one can put an upper limit on the overall magnitude
of a flat spectrum of perturbations at horizon crossing and from this an upper
limit on all higher multipoles. At the 90\% confidence  level an upper limit
of $a_2^2\approx 2\times 10^{-10}$ has been derived \cite{AbbSch}.
Equivalently, fitting observed clustering to a primordial fluctuations spectrum
\cite{bond} one can predict a value of $a_2^2$. Such an analysis yields best
fit values in the range $a_2^2\approx (1.9 -9.9)\times 10^{-11}$.
While these estimates are probably consistent with the COBE observation, they
also suggest that a major fraction of the observed anisotropy may be due to
gravitational waves.  Note that \cite{AbbWis2,AbbSch} both the gravitational
wave induced anisotropy and the density fluctuation induced anisotropy yield
similar distributions of moments at least up to $l\approx 10$.  Thus
measurements of the correlation function cannot easily serve to distinguish
between these possibilities at this stage.

What scale of inflation, M, in GeV, do the above limits correspond to?  From
(\ref{eqn:limits}) we find, at the 95\% confidence level,
$1.5\times 10^{16}\mbox{GeV}<M<5.2\times 10^{16}$GeV, with the best fit value
$2.9 \times 10^{16}$GeV.
On the other hand using data from precision electroweak measurements at LEP on
the strong and weak coupling constants one finds, for minimal $SU(5)$ SUSY
models with SUSY breaking between $M_Z$ and $1$TeV, that coupling constant
unification can occur at a single GUT scale,$M_X$ in the range
$M_X \approx (1-3.6) \times 10^{16}$GeV \cite{wilczek}.  Unfortunately there
are no explicit compelling SUSY or GUT inflationary scenarios with which one
can compare, but generically, unless there is fine tuning, or hierarchies, in
a GUT scenario  $V_0\approx \kappa M^4$ where $\kappa \approx .01-1$ (for
example in a Coleman-Weinberg $SU(5)$ model $\kappa=9/32\pi^2$).
Thus the energy scale of inflation consistent with the observed quadrupole
anisotropy coming from gravitational waves can coincide with the estimated
scale of SUSY Grand Unification as inferred from extrapolation of low energy
couplings.  We find this possibility both plausible and exciting.  At the very
least it is quite promising that COBE through the quadrupole anisotropy is
sensitive to gravitational waves from inflation at interesting scales.

Since both density perturbations and gravitational wave anisotropies resulting
from inflation result in a flat spectrum for the CMB anisotropy, with a great
similarity in all multipoles up to at least $l=9$, it will be difficult from
CMB measurements alone to verify whether or not the observed signal is due
gravitational waves.   How might one hope then to distinguish between these
possibilities?  The simplest way would be to probe for evidence of a flat
spectrum of gravitational waves in regions of smaller wavenumber.   At present,
the most sensitive gravitational wave detector at shorter wavelengths (periods
of $O$(years)) is also astrophysical in origin, and is based on timing
measurements of millisecond pulsars \cite{rees,kragw,taylor}.
On smaller wavelengths still terrestrial probes, such as the proposed LIGO
gravity wave detector \cite{thorne}, are currently envisaged.

The sensitivity of all such detectors is based on the mean energy density per
logarithmic frequency interval in gravitational waves.  For waves which come
inside the horizon during the matter dominated era we can utilize
(\ref{eqn:spectral-amp}) and (\ref{eqn:rv-norm}).  Averaging over many
wavelengths/periods, summing over helicities,  and also taking the stochastic
average, we find
\begin{equation}
  k {d\rho_g\over dk} = {k_{phys}^2\over 2G}
  \ A^2(k)\overline{\left[{3j_1(k\tau)\over k\tau}\right]^2}
\end{equation}
where $k_{phys}=k/R(\tau)$.  For wavelengths much smaller than the horizon
($k\tau\gg 2\pi$) this goes as $R^{-4}(\tau)$ as expected.  The time evolution
factor $\overline{(3j_1(k\tau)/k\tau)^2}$ is crucial, and in fact implies that
the energy density in gravitational waves also redshifts considerably as it
comes inside the horizon.  Thus the energy density at horizon crossing is
considerably smaller than the asymptotic value when the wave is well outside
the horizon, a fact which has not been stressed before to our knowledge.
In any case dividing by the critical density today we find for waves just
coming inside the horizon,
\begin{equation}
(\Omega_g)_{hc}\approx \left\{ \begin{array}{cccc}
 16v/9    &=& {2/3\pi}   ({H_{infl}/ {M_{Pl}}})^2   & RD \\
  v/\pi^2 &=& {3/8\pi^2} ({H_{infl}/ {M_{Pl}}})^2   & MD
\end{array} \right.
\end{equation}
The result at horizon crossing in a radiation dominated epoch was calculated
using the appropriate Bogolubov coefficients.  This results in the factor of
$3j_1(k\tau)/ k\tau$ above being changed to $j_0(k\tau)$. Waves which come
inside the horizon during the radiation dominated era will redshift with one
extra power of $R$ compared to matter during the matter dominated era.
Thus their contribution to $\Omega$ today will be suppressed compared to their
contribution at horizon crossing by the factor
$\rho_{rad}/\rho_c=4\times 10^{-5} h^{-2}$,
where the Hubble constant today is $100h$km/sec/Mpc. As a result, we find
that such waves today, taking $v<3.7\times 10^{-10}$, form a stochastic
background with $\Omega_g<2.6\times 10^{-14}h^{-2}<10^{-13};\ (h>0.5)$.

The waves for which the millisecond pulsar timing and future interferometer
measurements are sensitive entered the horizon during the radiation dominated
era.  The present limit, at the 68{\%} confidence level, from pulsar timing
data is $\Omega_g<9\times 10^{-8}$ \cite{taylor}.  This limit can improve in
principle with the measuring time to the fourth power \cite{rees,kragw} but,
even in the most optimistic case, dedicated observations with many pulsars,
and better clocks, over a period of perhaps a century would be required to
uncover such a signal.  The expected energy density is also about 2 orders of
magnitude below the optimum projected capabilities of future terrestrial
detectors.

Thus, prospects look grim, without some significant technological advances, for
detecting such a gravitational wave background directly anywhere but in the
microwave background signal.  Barring a very refined measurement of high
multipoles in the CMB anisotropy we may have to await confirmation at
accelerators, or perhaps proton decay detectors, before we can determine
whether COBE has discovered the first evidence for GUTs, supersymmetry, and
at the very least, gravitational waves.

\vskip .1in

\noindent We thank Mark Wise and Vince Moncrief for very helpful discussions.

\newpage

\noindent{\bf{Figure Captions}}

\vskip .1in

\noindent Figure 1:  The distribution for the scale of inflation
$v=(V_0/M_Pl^{4})$ as determined by Monte Carlo, using the COBE measurements
and assuming the observed quadrupole anisotropy is due to gravitational waves.

\end{document}